# On Formal Verification of Arithmetic-Based Cryptographic Primitives[*]

David Nowak

Research Center for Information Security, AIST, Japan

**Abstract.** Cryptographic primitives are fundamental for information security: they are used as basic components for cryptographic protocols or public-key cryptosystems. In many cases, their security proofs consist in showing that they are reducible to computationally hard problems. Those reductions can be subtle and tedious, and thus not easily checkable. On top of the proof assistant Coq, we had implemented in previous work a toolbox for writing and checking game-based security proofs of cryptographic primitives. In this paper we describe its extension with number-theoretic capabilities so that it is now possible to write and check arithmetic-based cryptographic primitives in our toolbox. We illustrate our work by machine checking the game-based proofs of unpredictability of the pseudo-random bit generator of Blum, Blum and Shub, and semantic security of the public-key cryptographic scheme of Goldwasser and Micali.

## 1 Introduction

Cryptographic primitives are fundamental components for information security. In many cases, their security proofs consist in showing that they are reducible to computationally hard problems. Those reductions can be subtle and tedious, and thus not easily checkable. Bellare and Rogaway even claim in [4] that:

> "many proofs in cryptography have become essentially unverifiable. Our field may be approaching a crisis of rigor."

As a remedy, they, and also Shoup [16], advocate game-based security proofs. This is a methodology for writing proofs which makes them easier to read and check. Halevi goes further by advocating the need for a software which can deal with the mundane parts of writing and checking game-based proofs [10].

In the game-based approach, a security property is modeled as a probabilistic program which implements a game to be solved by the attacker. The attacker itself is modeled as an external probabilistic procedure interfaced with the game. The goal is then to prove that any attacker has at most a negligible advantage over a random player. An attacker is assumed to be efficient i.e., it is modeled as a probabilistic polynomial-time (PPT) algorithm.

*Related Work.* There are tools such as ProVerif [5], CryptoVerif [6] or the prototype implementation of [12] which can make automatic proofs of cryptographic protocols or generic cryptographic schemes. However those tools assume that some secure cryptographic primitives are given. The security of those primitives cannot be proved automatically. Nevertheless their security proofs can be checked by a computer.

The game-based proof of the PRP/PRF switching lemma has been formalized in the proof assistant Coq [1]. Although it is not by itself a cryptographic primitive, this lemma is fundamental in proving security of some cryptographic schemes. The proof has been made in the random oracle model. The machine formalization in [1] is a so-called *deep embedding*: games are syntactic objects; and game transformations are syntactic manipulations which can be automated in the language of the proof assistant. The main advantages of this approach are that one can prove completeness of decision procedures, if any, and get smaller proof terms. However those advantages are not exploited in [1]. Moreover this is at the cost of developing a huge machinery for syntactic manipulations. Two other deep embeddings are currently being developed [2,3].

---

[*] This paper is an extended version of [14].

*Previous Work.* In cryptographers' papers, the formal semantics of games is either left implicit or, at best, informally explained in English. It is not enough for machine formalization. In previous work, we have (1) proposed a formal semantics for games, (2) implemented it in the proof assistant Coq, and (3) used it to prove the semantic security of the ElGamal and Hashed ElGamal public-key cryptographic schemes [13]. Our machine formalization is a so-called *shallow embedding*: games are probability distributions (as advocated by Shoup [16]). Game transformations can still be automated by going through the metalanguage of the proof assistant. Compared to [1], We have been very careful in making our design choices such that our implementation remains light. This is an important design issue in formal verification because formal proofs grow quickly in size when one tackles real-world use-cases.

Our toolbox comes in two layers. The first layer extends the standard library of Coq with mathematical notions and their properties that are fundamental in cryptography but not available in the standard library of Coq. This consists of a library for probability distributions, bitstrings, and a small library for elementary group theory. On top of this, the second layer consists of formal versions of security definitions and hard problems, and basic game transformations which can be composed to reduce the security of cryptographic primitives to hard problems.

By using our toolbox, one is forced to exhibit all the steps in his or her game-based proof and thus cannot hide assumptions or make proofs by intimidation such as "*Trivial*" or "*The reader may easily supply the details*". In spite of the required level of detail, proofs remain human readable and human checkable.

*Our contributions.* We have extended our toolbox in order to be able to deal with cryptographic primitives based on number theory. For that purpose we have added to the first layer a library of definitions and lemmas for integers modulo $n$. In particular we have formalized the notions of Legendre and Jacobi symbol, Blum primes, and their properties which are of fundamental use in cryptography. This is already by itself a contribution to the theorem proving community.[1] We have also considerably extended and generalized our library on elementary group theory.[2]

Then we have used our extension to the first layer in adding to the second layer the security notion of unpredictability, the quadratic residuosity assumption, and number-theoretic game transformations.

Finally, we have used our extensions to machine check the proof of unpredictability of the pseudo-random bit generator of Blum, Blum and Shub [7]. We have also machine checked the proof of semantic security of the public-key cryptographic scheme of Goldwasser and Micali [8]. Our security proofs of those two primitives are based on the intractability of the quadratic residuosity problem. To the best of our knowledge, this is the first time that security proofs of cryptographic primitives based on number theory are machine checked. This is also the first time that a proof of unpredictability is machine checked. None of the above mentioned related work could be used in their current state to formalize such proofs because they are missing components for number theory.

---

[1] Tools such as *Mathematica* can deal with formal computations involving Legendre and Jacobi symbols, but cannot be used to make formal proofs. In particular they do not allow reasoning by induction.

[2] There are more advanced library on group theory, such as [9], but none of them are available with the version of Coq that we are using (8.2beta3).

*Outline.* We introduce the proof assistant Coq in Sect. 2. In Sect. 3, we give formal meaning to games in terms of distributions. We give in Sect. 4 the number-theoretic facts that we use Sect. 5. In Sect. 5.1 we apply our work to the proof of unpredictability of the Blum-Blum-Shub generator, and in Sect. 5.2 we apply it to the proof of semantic security of the Goldwasser-Micali scheme. Finally, we briefly describe our implementation in Sect. 6 before concluding in Sect. 7.

## 2 The Coq proof assistant

Coq is a proof assistant developed at INRIA since 1984.[3] It is based on a kernel which takes a mathematical statement $S$ and proof term $p$ as input and check whether $p$ is a correct proof of $S$. On top of this kernel there are: a tactic language which allows to build proof terms in an incremental way; and decision procedures for decidable fragments such as Presburger arithmetic or propositional logic.

Coq is goal-directed. This means that if we are trying to prove that a formula $Q$ (the goal) is true, and we have an already proved theorem stating that $P_1$ & $P_2$ implies $Q$, then we can apply this theorem. Coq will replace the goal $Q$ by two subgoals $P_1$ and $P_2$. Proofs by induction are also possible. We proceed this way until we finally reach goals that are either axioms or are true by definition. On the way, Coq builds incrementally a proof term to be checked later by the kernel.

The kernel is the only critical part: if a bug outside the kernel causes a wrong proof term to be built, it will be rejected by the kernel.

In order to be closer to mathematical practice, Coq also provides mechanisms for introducing notations, or for inferring implicit parameters and subset coercions. It also comes with a standard library of definitions and lemmas, for instance on elementary arithmetic, analysis or polymorphic lists.

## 3 Games

We denote a game by a finite probability distribution (from now on, we will abbreviate this term as *distribution*). A distribution $\delta$ over a set $S$ is defined as a finite multiset[4] of ordered pairs from $S \times \mathbb{R}$ such that $\sum_{(a,p)\in\delta} p = 1$. We use the symbols $\{\!|$ and $|\!\}$ as delimiters of mutisets not to confuse them with sets. We write $p \cdot \{\!|(a_1, p_1), \ldots, (a_n, p_n)|\!\}$ for the multiset $\{\!|(a_1, p \cdot p_1), \ldots, (a_n, p \cdot p_n)|\!\}$.

For convenience, we introduce some notations (cf. Fig. 1) for writing distributions so that they will look like probabilistic programs:

– We write **return** $a$ for the distribution with only the element $a$ of probability 1.
– We write $x \Leftarrow \delta;\ \varphi(x)$ for the distribution built by picking at random a value $x$ according to the distribution $\delta$ and then computing the distribution $\varphi(x)$.
– We write $x \xleftarrow{R} \{a_1, \ldots, a_n\};\ \varphi(x)$ for the distribution built by picking at random a value $x$ in the set $\{a_1, \ldots, a_n\}$ and then computing the distribution $\varphi(x)$.
– We abbreviate this last case by $x \leftarrow a;\ \varphi(x)$ when the set is a singleton $\{a\}$.

---

[3] http://coq.inria.fr/
[4] A multiset (a.k.a. a bag) is a generalization of a set: a member of a multiset may be member more that once. For example, the multisets $\{\!|1, 2, 2|\!\}$ and $\{\!|1, 2|\!\}$ are different; and the union of $\{\!|1, 2, 2, 3|\!\}$ and $\{\!|1, 4, 4|\!\}$ is equal to $\{\!|1, 1, 2, 2, 3, 4, 4|\!\}$.

$$\textbf{return } a \;=\; \{|(a,1)|\} \qquad\qquad \begin{array}{l} x \Leftarrow \delta; \\ \varphi(x) \end{array} \;=\; \bigcup_{(a,p)\in\delta} p \cdot \varphi(a)$$

$$\begin{array}{l} x \xleftarrow{R} \{a_1,\ldots,a_n\}; \\ \varphi(x) \end{array} \;=\; \begin{array}{l} x \Leftarrow \{|(a_1,\tfrac{1}{n}),\ldots,(a_n,\tfrac{1}{n})|\}; \\ \varphi(x) \end{array}$$

$$\begin{array}{l} x \leftarrow a; \\ \varphi(x) \end{array} \;=\; \begin{array}{l} x \xleftarrow{R} \{a\}; \\ \varphi(x) \end{array}$$

**Fig. 1.** Notations for distributions

$$\begin{array}{l} x \leftarrow a; \\ \varphi(x) \end{array} = \varphi(a) \qquad \begin{array}{l} x \Leftarrow \delta; \\ \textbf{return } x \end{array} = \delta \qquad \begin{array}{l} y \Leftarrow ( \\ \quad x \Leftarrow \delta; \\ \quad \varphi(x) \\ ); \\ \psi(y) \end{array} = \begin{array}{l} x \Leftarrow \delta; \\ y \Leftarrow \varphi(x); \\ \psi(y) \end{array}$$

**Fig. 2.** Monad laws

Distributions have a monadic structure [15] and thus satisfy the monad laws (cf. Fig. 2). Those laws state that our notations for distributions behave well. The first one simply states that the occurences of a deterministically assigned variable $x$ can be replaced by their definition (constant propagation). The second one is a kind of $\eta$-reduction. The third one allows to simplify nested sequences.

We write $\Pr\left(P(\delta)\right)$ for the probability that $P$ holds of an element picked at random in the distribution $\delta$. Its value is

$$\sum_{(a,p)\in\delta \text{ s.t. } P(a)} p$$

We define a notion of indistinguishability for distributions. Security definitions, hard problems and game transformations will all be defined with this relation. Two distributions $\delta_1$ and $\delta_2$ are indistinguishable modulo $\epsilon$ w.r.t. a predicate $P$, written $\delta_1 \equiv^P_\epsilon \delta_2$, iff:

$$\left| \Pr\left(P(\delta_1)\right) - \Pr\left(P(\delta_2)\right) \right| \leq \epsilon$$

$\equiv^P_\epsilon$ is reflexive and symmetric. If $\delta_1 \equiv^P_\epsilon \delta_2$ and $\delta_2 \equiv^P_{\epsilon'} \delta_3$, then $\delta_1 \equiv^P_{\epsilon+\epsilon'} \delta_3$. If $\delta_1 \equiv^P_\epsilon \delta_2$ and $\epsilon \leq \epsilon'$, then $\delta_1 \equiv^P_{\epsilon'} \delta_2$. We write $\equiv_\epsilon$ instead of $\equiv^P_\epsilon$ when $P$ is the predicate on booleans such that $P(b)$ holds iff $b$ is equal to *true*.

**Lemma 3.1.** *If $f : S \to T$ is a bijection then, for all $\varphi$ and $P$,*

$$\begin{array}{l} x \xleftarrow{R} S; \\ \varphi(f(x)) \end{array} \equiv^P_0 \begin{array}{l} x \xleftarrow{R} T; \\ \varphi(x) \end{array}$$

*It is also true when $f$ is a surjective $N$-to-one function.*

This kind of transformation of one game into another one is at the crux of the security proofs we are dealing with.

## 4 Some elementary number theory

Let $n$ be a positive number. We write $\mathbb{Z}_n$ for the set of integers modulo $n$. The multiplicative group of $\mathbb{Z}_n$ is written $\mathbb{Z}_n^*$ and consists of the subset of integers modulo $n$ which are coprime with $n$. An integer $x \in \mathbb{Z}_n^*$ is a quadratic residue modulo $n$ iff there exists a $y \in \mathbb{Z}_n^*$ such that $y^2 \equiv x \pmod{n}$. Such a $y$ is called a square root of $x$ modulo $n$. We write $QR_n$ for the set of quadratic residues modulo $n$, and $QNR_n$ for its complement i.e., the set of quadratic non-residues modulo $n$. We write $\mathbb{Z}_n^*(+1)$ (respectively, $QNR_n(+1)$) for the subset of integers in $\mathbb{Z}_n^*$ (respectively, $QNR_n$) with Jacobi symbol equal to 1.

The quadratic residuosity problem is the following: given an odd composite integer $n$, decide whether or not an $x \in \mathbb{Z}_n^*$ is a quadratic residue modulo $n$.

Let $n$ be the product of two distinct odd primes $p$ and $q$. The quadratic residuosity assumption (QRA) states that the above problem is intractable. In our framework, this can be stated as:

**Assumption 4.1 (QRA).** *For every attacker $A'$, there exists a negligible $\epsilon$ such that*

$$
\begin{array}{l}
x \xleftarrow{R} \mathbb{Z}_n^*(+1); \\
\widehat{b} \Leftarrow A'(n, x); \\
b \leftarrow \widehat{b} = \mathsf{qr}(x); \\
\textbf{return } b
\end{array}
\equiv_\epsilon
\begin{array}{l}
b \xleftarrow{R} \{true, false\}; \\
\textbf{return } b
\end{array}
$$

In the left-side game, an $x$ is picked at random in the set $\mathbb{Z}_n^*(+1)$; this $x$ is passed with $n$ to the attacker $A'$; the attacker returns its guess $\widehat{b}$ for the quadratic residuosity (modulo $n$) of $x$; this guess is compared with the true quadratic residuosity (modulo $n$) $\mathsf{qr}(x)$ of $x$; and the result $b$ of this comparison is returned. In the rigth-side game, the result is random. QRA states that the advantage $\epsilon$ of any attacker over a random player is negligible.

Note that the fact that $A'$ is a randomized algorithm is modeled by the attacker returning a distribution in which $\widehat{b}$ is picked.

In the security proofs, we will need the following well-known mathematical facts (remember that $n$ is the product of two distinct odd primes $p$ and $q$):

**Fact I.** The function which maps an $x \in \mathbb{Z}_n^*$ to $x^2 \in QR_n$ is a surjective four-to-one function.

**Fact II.** For any $y \in QNR_n(+1)$, the function which maps an $x \in QR_n$ to $y \cdot x \in QNR_n(+1)$ is a bijection.

**Fact III.** $|QR_n| = |QNR_n(+1)|$

**Fact IV.** $\mathbb{Z}_n^*(+1) = QR_n \cup QNR_n(+1)$

Let $n$ be a Blum integer i.e., the product of two distinct prime numbers $p$ and $q$, each congruent to 3 modulo 4. In this case, any $x \in QR_n$ has a unique square root in $QR_n$ which we denote by $\sqrt{x}$ and is called the principal square root of $x$. And we get the following additional facts [7]:

**Fact V.** The function which maps an $x \in QR_n$ to $x^2 \in QR_n$ is a permutation.

**Fact VI.** The function which maps an $x \in \mathbb{Z}_n^*(+1)$ to $x^2 \in QR_n$ is a surjective two-to-one function.

**Fact VII.** For all $x \in QR_n, \quad \sqrt{x^2} = x$

**Fact VIII.** For all $x \in \mathbb{Z}_n^*(+1), \quad x \in QR_n \Leftrightarrow \mathsf{parity}(x) = \mathsf{parity}(\sqrt{x^2})$

## 5 Applications

In this section, we apply our work to the proofs of unpredictability of the Blum-Blum-Shub generator, and to the proof of semantic security of the Goldwasser-Micali scheme.

$$\begin{aligned}&\mathsf{bbs}(len \in \mathbb{N}, seed \in \mathbb{Z}_n^*) = \\ &\quad \mathsf{bbs\_rec}(len, seed^2)\end{aligned} \qquad \begin{aligned}&\mathsf{bbs\_rec}(len \in \mathbb{N}, x \in QR_n) = \\ &\quad \textbf{match } len \textbf{ with} \\ &\quad | 0 \Rightarrow [] \\ &\quad | len' + 1 \Rightarrow \mathsf{parity}(x) :: \mathsf{bbs\_rec}(len', x^2) \\ &\quad \textbf{end}\end{aligned}$$

**Fig. 3.** The Blum-Blum-Shub generator

### 5.1 The Blum-Blum-Shub pseudorandom bit generator

The security of many cryptographic systems depends upon a cryptographic primitive for the generation of unpredictable sequences of bits. They are used to generate keys, nonces or salts. Ideally, those sequences of bits should be random, that is, generated by successive flips of a fair coin. In practice, one uses a pseudorandom bit generator (PRBG) which, given a short seed, generates a long sequence of bits that appears random. For the purpose of simulation, one only requires of a PRBG that it passes certain statistical tests (cf. Chapter 3 of [11]). This is not enough for cryptography. A PRBG is cryptographically secure iff it passes all polynomial-time statistical tests: roughly speaking, no polynomial-time algorithm can distinguish between an output sequence of the generator and a truly random sequence.

In [7], the Blum-Blum-Shub generator (BBS) is proved left-unpredictable (under the quadratic residuosity assumption). It was proved by Yao in [18] that this is equivalent to stating that BBS passes all polynomial-time statistical tests. It is shown in [17] that BBS is still secure under the weaker assumption that $n$ is hard to factorize. The same authors also show that, for sufficiently large $n$, more than one bit can be extracted at each iteration of the algorithm. However, in this paper, we stick to the original proof of [7].

Let $n = p \cdot q$ be a Blum integer. The BBS generator is defined by the function bbs given in Fig. 3 which takes as input a length and a seed, and returns a pseudorandom sequence of bits of the required length.

In our framework, one can state the left-unpredictability of bbs by the following definition.

**Definition 5.1 (Left-unpredictability).** bbs *is left-unpredictable iff for all length len, for every attacker A, there exists a negligible $\epsilon$ such that*

$$\begin{aligned}&seed \xleftarrow{R} \mathbb{Z}_n^*; \\ &[b_0, \ldots, b_{len}] \leftarrow \mathsf{bbs}(len+1, seed); \\ &\widehat{b_0} \Leftarrow A([b_1, \ldots, b_{len}]); \qquad \equiv_\epsilon \quad \begin{aligned}&b \xleftarrow{R} \{true, false\}; \\ &\textbf{return } b\end{aligned} \\ &b \leftarrow \widehat{b_0} = b_0; \\ &\textbf{return } b\end{aligned}$$

In the left-side game, a *seed* is picked at random in the set $\mathbb{Z}_n^*$; the function bbs is then used to compute a pseudorandom sequence of bits $[b_0, \ldots, b_{len}]$ of length $len + 1$; this sequence minus its first bit $b_0$ is passed to the attacker $A$; the attacker returns its guess $\widehat{b_0}$ for the value of the bit $b_0$; this guess is compared with $b_0$; and the result $b$ of this comparison is returned. bbs is left-unpredictable if the advantage $\epsilon$ of any attacker over a random player is negligible.

Before proving that bbs is left-unpredictable, we show that it can be reduced to the problem of finding the parity of a random quadratic residue modulo $n$.

**Lemma 5.2.** *If, for every attacker $A'$, there exists a negligible $\epsilon$ such that*

$$\begin{aligned}&x \xleftarrow{R} QR_n; \\ &\widehat{b} \Leftarrow A'(n, x); \qquad \equiv_\epsilon \quad \begin{aligned}&b \xleftarrow{R} \{true, false\}; \\ &\textbf{return } b\end{aligned} \\ &b \leftarrow \widehat{b} = \mathsf{parity}(\sqrt{x}); \\ &\textbf{return } b\end{aligned}$$

*then* bbs *is left-unpredictable.*

In the left-side game, $x$ is picked at random in the set $QR_n$; this $x$ is passed with $n$ to the attacker $A'$; the attacker returns its guess $\widehat{b}$ for the parity of $\sqrt{x}$; this guess is compared with the true parity of $\sqrt{x}$; and the result $b$ of this comparison is returned. The above lemma states that if the advantage $\epsilon$ of any attacker over a random player is negligible, then bbs is left-unpredictable.

*Proof (of Lemma 5.2).* We proceed by rewriting the left-side game of the left-unpredictability specification (Def. 5.1).

**BBS1.** We unfold the definition of bbs:

$$seed \xleftarrow{R} \mathbb{Z}_n^*;$$
$$[b_0, \ldots, b_{len}] \leftarrow \mathsf{bbs\_rec}(len + 1, seed^2);$$
$$\widehat{b_0} \Leftarrow A([b_1, \ldots, b_{len}]);$$
$$b \leftarrow \widehat{b_0} = b_0;$$
$$\textbf{return } b$$

**BBS2.** Because of Fact I, we can rewrite the game as:

$$x \xleftarrow{R} QR_n;$$
$$[b_0, \ldots, b_{len}] \leftarrow \mathsf{bbs\_rec}(len + 1, x);$$
$$\widehat{b_0} \Leftarrow A([b_1, \ldots, b_{len}]);$$
$$b \leftarrow \widehat{b_0} = b_0;$$
$$\textbf{return } b$$

**BBS3.** $x$ is a quadratic residue, we can thus replace $x$ with $\sqrt{x^2}$ (according to Fact VII).

$$x \xleftarrow{R} QR_n;$$
$$[b_0, \ldots, b_{len}] \leftarrow \mathsf{bbs\_rec}(len + 1, \sqrt{x^2});$$
$$\widehat{b_0} \Leftarrow A([b_1, \ldots, b_{len}]);$$
$$b \leftarrow \widehat{b_0} = b_0;$$
$$\textbf{return } b$$

**BBS4.** Because of Fact V, we can rewrite the game as:

$$x \xleftarrow{R} QR_n;$$
$$([b_0, \ldots, b_{len}]) \leftarrow \mathsf{bbs\_rec}(len + 1, \sqrt{x});$$
$$\widehat{b_0} \Leftarrow A([b_1, \ldots, b_{len}]);$$
$$b \leftarrow \widehat{b_0} = b_0;$$
$$\textbf{return } b$$

**BBS5.** By unfolding one step of bbs_rec, we get:

$$x \xleftarrow{R} QR_n;$$
$$[b_1, \ldots, b_{len}] \leftarrow \mathsf{bbs\_rec}(len, x);$$
$$\widehat{b_0} \Leftarrow A([b_1, \ldots, b_{len}]);$$
$$b \leftarrow \widehat{b_0} = \mathsf{parity}(\sqrt{x});$$
$$\textbf{return } b$$

**BBS6.** We have reduced the game to the left-side one of the hypothesis where the attacker $A'(n, x)$ is instantiated by:

$$[b_1, \ldots, b_{len}] \leftarrow \mathsf{bbs\_rec}(len, x);$$
$$\widehat{b_0} \Leftarrow A([b_1, \ldots, b_{len}]);$$
$$\textbf{return } \widehat{b_0} \quad \square$$

Using the above lemma, we can now prove that bbs is left-unpredictable.

**Theorem 5.3.** bbs *is left-unpredictable (under the quadratic residuosity assumption).*

*Proof.* By Lemma 5.2, we only need to prove that for every attacker $A$:

$$\begin{array}{l} x \xleftarrow{R} QR_n; \\ \widehat{b} \Leftarrow A(n, x); \\ b \leftarrow \widehat{b} = \mathsf{parity}(\sqrt{x}); \\ \textbf{return } b \end{array} \equiv_\epsilon \begin{array}{l} b \xleftarrow{R} \{true, false\}; \\ \textbf{return } b \end{array}$$

We proceed by rewriting the left-side game.

**BBS7.** Because of Fact VI, we can rewrite the game as:

$$x \xleftarrow{R} \mathbb{Z}_n^*(+1);$$
$$\widehat{b} \Leftarrow A(n, x^2);$$
$$b \leftarrow \widehat{b} = \mathsf{parity}(\sqrt{x^2});$$
$$\textbf{return } b$$

**BBS8.** By Fact VIII, we can replace the equality test $\widehat{b} = \mathsf{parity}(\sqrt{x^2})$ by $\widehat{b} \oplus \mathsf{parity}(x) \oplus 1 = \mathsf{qr}(x)$ (where $\oplus$ is the notation for the exclusive-or XOR).

$$x \xleftarrow{R} \mathbb{Z}_n^*(+1);$$
$$\widehat{b} \Leftarrow A(n, x^2);$$
$$b \leftarrow \widehat{b} \oplus \mathsf{parity}(x) \oplus 1 = \mathsf{qr}(x);$$
$$\textbf{return } b$$

**BBS9.** We have reduced the game to the left-sided one of QRA (Assumption 4.1) where the attacker $A'(n, x)$ is instantiated by:

$$\widehat{b} \Leftarrow A(n, x^2);$$
$$\textbf{return } \widehat{b} \oplus \mathsf{parity}(x) \oplus 1 \quad \square$$

### 5.2 The Goldwasser-Micali public-key cryptographic scheme

The Goldwasser-Micali public-key cryptographic scheme (GM) was the first probabilistic one which was provably secure. More precisely it is semantically secure under the quadratic residuosity assumption [8]. For defining $GM$, we need a number $n$ which is the product of two distinct prime numbers $p$ and $q$, and a $y \in QNR_n(+1)$. It is then defined by the three functions given in Fig. 4 where $\left(\frac{c}{p}\right)$ denotes the Legendre symbol of $c$.

In our framework, one can state the semantic security of the above scheme by the following definition.

$$
\begin{array}{ll}
\text{keygen}() = & \text{encrypt}((n,y) \in \mathbb{Z} \times \mathbb{Z}_n^*, b \in \{0,1\}) = \\
\quad pk \leftarrow (n,y); & \quad x \stackrel{R}{\leftarrow} \mathbb{Z}_n^*; \\
\quad sk \leftarrow (p,q); & \quad c \leftarrow (\text{if } b=1 \text{ then } y \cdot x^2 \text{ else } x^2); \\
\quad \textbf{return } (pk, sk) & \quad \textbf{return } c
\end{array}
$$

$$
\begin{aligned}
&\text{decrypt}((p,q) \in \mathbb{Z} \times \mathbb{Z}, c \in \mathbb{Z}_n^*) = \\
&\quad e \leftarrow \left(\frac{c}{p}\right); \\
&\quad m \leftarrow (\text{if } e=1 \text{ then } 0 \text{ else } 1); \\
&\quad \textbf{return } m
\end{aligned}
$$

**Fig. 4.** The Goldwasser-Micali scheme

**Definition 5.4.** *GM is semantically secure iff, for every attacker $(A_1, A_2)$, there exists a negligible $\epsilon$ such that*

$$
\begin{array}{l}
(pk, sk) \Leftarrow \text{keygen}(); \\
(m_1, m_2) \Leftarrow A_1(pk); \\
i \stackrel{R}{\leftarrow} \{1,2\}; \\
c \Leftarrow \text{encrypt}(pk, m_i); \\
\widehat{\imath} \Leftarrow A_2(pk, (m1, m2), c); \\
\textbf{return } \widehat{\imath} = i
\end{array}
\equiv_\epsilon
\begin{array}{l}
b \stackrel{R}{\leftarrow} \{true, false\}; \\
\textbf{return } b
\end{array}
$$

In the left-side game, a pair $(pk, sk)$ of public and secret keys is generated; the public key $pk$ is passed to the attacker $A_1$ which returns two messages $m_1$ and $m_2$; one of them is picked at random and encrypted with the secret key $sk$; the obtained cyphertext $c$ is then passed with the public key $pk$ and the pair of picked messages $(m_1, m_2)$ to the attacker $A_2$; the attacker returns its guess for the picked message; whether the attacker is right or not is returned as a result. A scheme is semantically secure if the advantage $\epsilon$ of any attacker over a random player is negligible.

**Theorem 5.5.** *The scheme of Goldwasser and Micali is semantically secure (under the quadratic residuosity assumption).*

*Proof.* We proceed by rewriting the left-side game of the semantic-security specification (Def. 5.4).

**GM1.** We unfold definitions of keygen and encrypt:

$$
\begin{array}{l}
(m_1, m_2) \Leftarrow A_1(n, y); \\
i \stackrel{R}{\leftarrow} \{1,2\}; \\
x \stackrel{R}{\leftarrow} \mathbb{Z}_n^*; \\
\widehat{\imath} \Leftarrow A_2((n,y), (m_1, m_2), \textbf{if } m_i = 1 \textbf{ then } y \cdot x^2 \textbf{ else } x^2); \\
\textbf{return } \widehat{\imath} = i
\end{array}
$$

**GM2.** Because of Fact I, we can rewrite the game as:

$$
\begin{array}{l}
(m_1, m_2) \Leftarrow A_1(n, y); \\
i \stackrel{R}{\leftarrow} \{1,2\}; \\
x \stackrel{R}{\leftarrow} QR_n; \\
\widehat{\imath} \Leftarrow A_2((n,y), (m_1, m_2), \textbf{if } m_i = 1 \textbf{ then } y \cdot x \textbf{ else } x); \\
\textbf{return } \widehat{\imath} = i
\end{array}
$$

**GM3.** Because of Fact II, we can rewrite the game as:

$$(m_1, m_2) \Leftarrow A_1(n, y);$$
$$i \xleftarrow{R} \{1, 2\};$$
$$x \xleftarrow{R} QR_n;$$
$$z \xleftarrow{R} QNR_n(+1);$$
$$\widehat{i} \Leftarrow A_2((n, y), (m_1, m_2), \text{if } m_i = 1 \text{ then } z \text{ else } x);$$
$$\textbf{return } \widehat{i} = i$$

Note that this transformation is only valid because the result of the game does not depend on the relation between $y \cdot x$ and $x$. Indeed, if $m_i = 1$ then only $y \cdot x$ is used while $x$ can be ignored, and vice versa.

Now we consider the different cases for the messages $m_1$ and $m_2$ chosen by the attacker.

(i) $(m_1, m_2) = (0, 0)$:

**GM4.** We can rewrite the game as:

$$x \xleftarrow{R} QR_n;$$
$$z \xleftarrow{R} QNR_n(+1);$$
$$\widehat{i} \Leftarrow A_2((n, y), (0, 0), x);$$
$$i \xleftarrow{R} \{1, 2\};$$
$$\textbf{return } \widehat{i} = i$$

$i$ can be picked randomly after the calls to the attacker. Therefore $\widehat{i}$ does not depend on $i$. Our goal is proved.

(ii) $(m_1, m_2) = (1, 1)$: This is similar to the previous case except that $A_2$ is given $z$ instead of $x$.

(iii) $(m_1, m_2) = (0, 1)$:

**GM5.** We can rewrite the game GM3 as:

$$i \xleftarrow{R} \{1, 2\};$$
$$\textbf{if } i = 1 \textbf{ then}$$
$$\quad x \xleftarrow{R} QR_n;$$
$$\quad \widehat{i} \Leftarrow A_2((n, y), (0, 1), x);$$
$$\quad \textbf{return } \widehat{i} = i$$
$$\textbf{else}$$
$$\quad z \xleftarrow{R} QNR_n(+1);$$
$$\quad \widehat{i} \Leftarrow A_2((n, y), (0, 1), z);$$
$$\quad \textbf{return } \widehat{i} = i$$

**GM6.** By definition of qr, we can rewrite the game as:

$$
\begin{aligned}
&i \xleftarrow{R} \{1,2\}; \\
&\textbf{if } i = 1 \textbf{ then} \\
&\quad x \xleftarrow{R} QR_n; \\
&\quad \widehat{\imath} \Leftarrow A_2((n,y),(0,1),x); \\
&\quad \widehat{b} \leftarrow \widehat{\imath} = 1; \\
&\quad \textbf{return } \widehat{b} = \mathsf{qr}(x) \\
&\textbf{else} \\
&\quad z \xleftarrow{R} QNR_n(+1); \\
&\quad \widehat{\imath} \Leftarrow A_2((n,y),(0,1),z); \\
&\quad \widehat{b} \leftarrow \widehat{\imath} = 1; \\
&\quad \textbf{return } \widehat{b} = \mathsf{qr}(z)
\end{aligned}
$$

**GM7.** Because of Fact III, we can rewrite the game as:

$$
\begin{aligned}
&x \xleftarrow{R} QR_n \cup QNR_n(+1); \\
&\widehat{\imath} \Leftarrow A_2((n,y),(0,1),x); \\
&\widehat{b} \leftarrow \widehat{\imath} = 1; \\
&\textbf{return } \widehat{b} = \mathsf{qr}(x)
\end{aligned}
$$

**GM8.** Because of Fact IV, we can rewrite the game as:

$$
\begin{aligned}
&x \xleftarrow{R} \mathbb{Z}_n^*(+1); \\
&\widehat{\imath} \Leftarrow A_2((n,y),(0,1),x); \\
&\widehat{b} \leftarrow \widehat{\imath} = 1; \\
&\textbf{return } \widehat{b} = \mathsf{qr}(x)
\end{aligned}
$$

**GM9.** We have reduced the game to the left-side one of QRA (Assumption 4.1) where the attacker $A'(n,x)$ is instantiated by:

$$
\begin{aligned}
&\widehat{\imath} \Leftarrow A_2((n,y),(0,1),x); \\
&\widehat{b} \leftarrow \widehat{\imath} = 1; \\
&\textbf{return } \widehat{b}
\end{aligned}
$$

(iv) $(m_1, m_2) = (1, 0)$: This case is similar to the previous one. □

## 6 Implementation

We have extended our toolbox with a module which contains number-theoretic lemmas on Legendre and Jacobi symbols, and on Blum integers. Based on those lemmas, we have proved arithmetic-based game transformations in the module dedicated to transformations. We have added the definition of unpredictability and the quadratic residuosity assumption in the appropriate modules. We have then used those extensions to make the formal security proofs of the Blum-Blum-Shub generator and the Goldwasser-Micali scheme.

*Future work.* Two standard mathematical results remain to be proved in the proof assistant Coq. The first one is Fermat's little theorem. Although there are proofs of this theorem in the contributions of Coq, they are not compatible with its standard library. The second one is the fact that if $p$ is a prime number then the group $\mathbb{Z}_p^*$ is cyclic. Although those theorems are orthogonal to our work, it would be nice to have them machine checked, if only for the sake of completeness. For the moment, we have added them as axioms.

We neither compute exact nor asymptotic running time. This is orthogonal to the verification of game transformations. In the examples we dealt with, the algorithm $A'$ we built from the attacker $A$ at the end of Lemma 5.2, Theorem 5.3 and Theorem 5.5 are trivially PPT and thus valid attackers. However this is not checked by the current implementation.

## 7 Conclusions

We have extended our toolbox with number-theoretic capabilities. It is thus now possible to use this toolbox for machine-checking game-based proofs of arithmetic-based cryptographic primitives. We have shown usability of our implementation by applying it to the proof of unpredictability of the Blum-Blum-Shub generator and the proof of semantic security of the Goldwasser-Micali scheme. This is the first time that a proof of unpredictability is machine checked. Machine formalization has forced us to make clear all details in those proofs that are usually either left to the reader or roughly explained in English. In spite of this level of details, we claim that our proofs remain human readable and are mechanically human checkable without appealing too much to intuition.

### Acknowledgements

We are grateful to Frédérique Oggier and Nicolas Perrin for fruitful discussions.